\documentclass[referee]{aa}
\usepackage{graphicx,amssymb}

\begin{document}

%Personal Macros
\def\etal{{\it et al.} }
\def\araa{{\it Ann.\ Rev.\ Astron.\ Ap.}}
\def\aplet{{\it Ap.\ Letters}}
\def\aj{{\it Astron.\ J.}}
\def\apj{{\it ApJ}}
\def\apjl{{\it ApJ\ Lett.}}
\def\apjs{{\it ApJ\ Suppl.}}
\def\aas{{\it Astron.\ Astrophys.\ Suppl.}}
\def\aa{{\it A\&A}}
\def\mnras{{\it MNRAS}}
\def\nature{{\it Nature}}
\def\pasa{{\it Proc.\ Astr.\ Soc.\ Aust.}}
\def\pasp{{\it P.\ A.\ S.\ P.}}
\def\pasj{{\it PASJ}}
\def\pre{{\it Preprint}}
\def\aph{{\it Astro-ph}}
\def\sovlet{{\it Sov. Astron. Lett.}}
\def\adspr{{\it Adv. Space. Res.}}
\def\expas{{\it Experimental Astron.}}
\def\ssr{{\it Space Sci. Rev.}}
\def\ar{{\it Astronomy Reports}}
\def\inpress{{\it in press}}
\def\inprep{{\it in prep.}}
\def\submit{{\it submitted}}

\def\ap{$\approx$ }
\def\mjysr{MJy/sr }
\def\inu{{I_{\nu}}}
\def\inufit{I_{\nu fit}}
\def\fnu{{F_{\nu}}}
\def\bnu{{B_{\nu}}}
\def\msol{{M$_{\odot}$}}
\def\mic{{\mu}m}
\def\cm2{$cm^{-2}$}

\title{Submillimeter dust emission of the M17 complex measured with PRONAOS} 

\author{X. Dupac\inst{1}, M. Giard\inst{1}, J.-P. Bernard\inst{1}$^,$\inst{2},
  N. Boudet\inst{1}, J.-M. Lamarre\inst{3}, C. M\'eny\inst{1},
  F. Pajot\inst{2}, \'E. Pointecouteau\inst{1},
  I. Ristorcelli\inst{1}, G. Serra\inst{1}, B. Stepnik\inst{2}, J.-P. Torre\inst{4}}
\institute{Centre d'\'Etude Spatiale des Rayonnements \\ 
9 av. du colonel Roche, BP4346, F-31028 Toulouse cedex 4, France
\and
Institut d'Astrophysique Spatiale \\
Campus d'Orsay B\^at. 121 \\
15, rue Cl\'emenceau, F-91405 Orsay cedex, France
\and
LERMA, Observatoire de Paris \\
61, Avenue de l'Observatoire, F-75014 Paris, France
%\and
%Tohoku University, Astronomical Institute, Sendai 980-8577, Japan
\and
Service d'A\'eronomie du CNRS \\
BP3, F-91371 Verri\`eres-le-Buisson cedex, France
}

\offprints{dupac@cesr.fr}

\authorrunning{Dupac \etal}
\titlerunning{Dust emission in M17}

\date{Received {15 January 2002} /Accepted {12 June 2002}}

\abstract{We map a 50$'$ x 30$'$ area in and around the M17 molecular complex
  with the French submillimeter balloon-borne telescope PRONAOS, in order to
  better understand the thermal emission of cosmic dust and the structure of
  the interstellar medium.
The PRONAOS-SPM instrument has an angular resolution of
about 3$'$, corresponding to a size of 2 pc at the distance of this complex,
and a high sensitivity up to 0.8 MJy/sr.
The observations are made in four wide submillimeter bands corresponding to effective wavelengths
of 200 $\mic$, 260 $\mic$, 360 $\mic$ and 580 $\mic$.
Using an improved map-making method for PRONAOS data, we map the M17
complex and faint condensations near the dense warm core.
We derive maps of both the dust temperature and the spectral index, which vary
over a wide range, from about 10 K to 100 K for the temperature and from about 1
  to 2.5 for the spectral index.
We show that these parameters are anticorrelated, the
  cold areas (10-20 K) having a spectral index around 2, whereas the warm
  areas have a spectral index between 1 and 1.5.
We discuss possible causes of this effect, and we propose an
explanation involving intrinsic variations of the grain properties.
Indeed, to match the observed spectra with two dust components having a spectral index equal to 2 leads to very large and unlikely amounts of cold dust.
We also give estimates of the column densities and masses of the studied
clumps.
Three cold clumps (14-17 K) could be gravitationally unstable.
}

\maketitle

\keywords{dust --- infrared: ISM: continuum --- ISM: clouds --- ISM: individual (M17)}

\section{Introduction\label{intro}}

The submillimeter domain is particularly suited to characterizing dust
properties in the interstellar medium.
Dust emission in this spectral range is mainly due to big grains at thermal
equilibrium (see, e.g., \cite{desert90}), whose emission is usually modelled
by the modified blackbody law, {\it i.e.} by the temperature and the spectral
index of the dust.
The temperature of a molecular cloud is a key parameter which controls
(with others) the structure and evolution of the clumps, and therefore, star formation.
Thus spectral imaging of molecular clouds can provide useful
information about their structure and evolution, especially if the dust emission
parameters can be properly derived on top of submillimeter intensities.
Mapping of star-forming molecular clouds, as well as other dusty regions, has
been performed by the PRONAOS balloon-borne experiment (PROgramme NAtional d'Observations
Submillim\'etriques, see \cite{serra01} or \cite{ristorcelli98}).
We present in this article the maps and analysis made from PRONAOS
observations of the M17 star-forming complex.

The Messier 17 Nebula (also called Omega, Horseshoe or Swan Nebula) is an
ionized region associated with a giant
star-forming molecular cloud located at about 2200
parsecs from us (\cite{chini80}) in the constellation of Sagittarius.
This nebula has the largest known ionization rate in the Galaxy for star
forming regions (see for example \cite{glushkov98}).
This large ionization rate is mainly due to the excitation of gas by young O-type stars.
The M17 molecular cloud has been mapped in carbon monoxide emission by Lada (1976), showing
the two most intense condensations in this cloud, usually called M17 North (N)
and M17 Southwest (SW).
This cloud is part of a giant molecular complex extending 170 pc to the
southwest, along the Sagittarius spiral arm (CO emission shown by \cite{elmegreen79}).
The interaction of this giant molecular cloud with the H II regions is
particularly visible in the M17 region, where an expanding shock front
interacts with the gas clouds, and is thought to have fragmented the original
molecular cloud (\cite{rainey87}).
The M17 SW cloud is the best studied region of the area, especially for the
photon-dominated region (PDR) near the boundary of the H II region (to the
northeast).
The first far infrared observations of M17 were made by Low \& Aumann (1970)
and Harper \& Low (1971).
M17 SW was mapped in the mid and far-infrared by Harper \etal (1976)
and Gatley \etal (1979).
Wilson \etal (1979) mapped the whole M17 cloud ({\it i.e.}
including M17 North) at 69 $\mic$.
The IRAS satellite (http://www.ipac.caltech.edu/ipac/iras/iras.html)
provided infrared maps of the whole sky, which showed the distribution of the
dust in the M17 complex, however without giving much information on the
faintest and coldest regions of the complex.
The recent JCMT measurement of Wilson \etal (1999) provided precise CO maps of
the M17 molecular cloud, as well as the measurement of Sekimoto \etal (1999)
on a larger area.
In this context, submillimeter mapping can provide useful
information about the dust properties, especially in cold regions, and
give independent estimations of the ISM masses in this region.
For these reasons, we have observed a large region including the
M17 cloud and fainter areas, with the multi-band photometric instrument
(SPM) of the PRONAOS balloon-borne experiment.
The maps are 50$'$ by 30$'$ (about 30 pc x 20 pc at the distance of M17) at angular resolutions between 2$'$ and 3.5$'$.
We present in Section 2 how the observations were made and the way we
processed the data to make maps.
Section 3 presents the resulting maps, and Section 4 presents a detailed
analysis of these maps.

\section{Observations and data processing\label{obs}}

PRONAOS (PROgramme NAtional d'Observations Submillim\'etriques) is a French
balloon-borne submillimeter experiment.
The 2 m telescope is described in detail in Buisson \& Duran (1990).
The focal plane instrument SPM (Syst\`eme Photom\'etrique Multibande, see \cite{lamarre94}) is
composed of a wobbling mirror, providing a beam switching on the sky with an
amplitude of about 6$'$ at 19.5 Hz, and four bolometers cooled at 0.3 K.
They measure the submillimeter flux in the spectral ranges 180-240 $\mic$, 240-340
$\mic$, 340-540 $\mic$ and 540-1200 $\mic$, with sensitivity to low
brightness gradients of about 4 \mjysr in band 1 and 0.8 \mjysr in band 4.
The effective wavelengths are 200, 260, 360 and 580 $\mic$, and the angular resolutions are
2$'$ in bands 1 and 2, 2.5$'$ in band 3 and 3.5$'$ in band 4.
Details about the instrument can be found in Ristorcelli \etal (1998).
The data which we present here were obtained during the second flight of PRONAOS in
september 1996, at Fort Sumner, New Mexico.

The usual reconstruction method from chopped PRONAOS data was EKH-like (\cite{emerson79})
with a scan by scan filtering in Fourier space (see \cite{sales91}).
Dupac \etal (2001) developed another method, based on direct linear
inversion on the whole map, using a Wiener matrix.
This map-making process takes into account the beam sizes and profiles, the
beam switching and the signal and noise properties to construct an optimal
map.

For these M17 data, we have used a slightly improved method, in which we
consider some noise not independent of the sky signal.
To assume a perfect independence between noise and signal is justified
for Cosmic Microwave Background maps, in which the intensity contrasts are low
(see map-making methods in \cite{dupac02}).
However, in the case of timelines having much contrast such as the PRONAOS ones,
the instrument response can hardly be considered as perfect.
Thus, we allowed the noise covariance matrix
to take into account noise proportionally correlated to the signal.
The fraction of the noise that appears to be correlated to the signal is determined by
iterations and tests of the reconstruction method.
We have compared PRONAOS M17 maps made with the method of Dupac \etal
(2001), {\it i.e.} assuming that the noise is independent of the signal, and the
ones made with the new method.
In the case of the independent-noise method, it appears that the found noise level
constrains the lowest signal that can be recovered.
The problem is that if the noise is stronger in intense areas, due to its correlation
with the signal, then the independent-noise method removes signal features lower than
this noise level.
With the new method, the problem is solved by allowing the noise not to be the
same in faint areas and in intense areas.
We found that the maps made with the new method are significantly better as they
reconstruct more faint areas.
The characteristics of this new method are otherwise common to the one
described in Dupac \etal (2001).

\section{Results\label{res}}

\begin{figure}[]
%\vbox to 10cm{
%\epsfxsize=14cm
\includegraphics[scale=1.3]{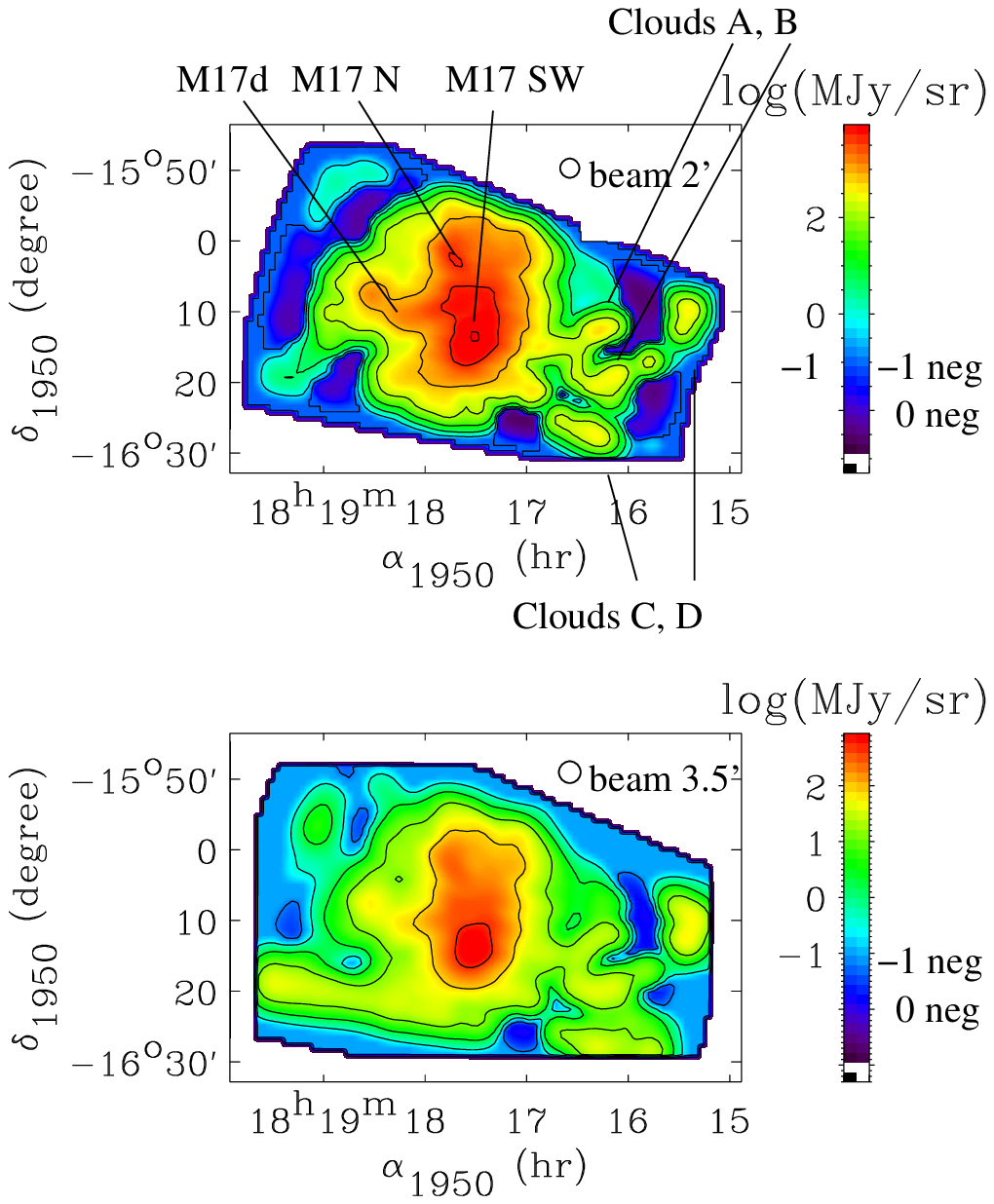}
\caption[]{PRONAOS intensity maps in band 1 (200 $\mic$, up) and band 4 (580 $\mic$,
  bottom).
The angular resolution is 2$'$ for the 200 $\mic$ map and 3.5$'$ for the 580
$\mic$ map.
The color scale is logarithmic, and displays the positive reconstructed intensity
until -1 in log, then the deeper blue and purple colors display the negative
noise features.
The noise level is about 4 \mjysr rms in band 1 and 0.8 \mjysr in band 4.
Due to the calibration uncertainty, the intensity accuracy is 5 \% (1
$\sigma$) relative between bands (8 \% absolute).
The non-observed areas are displayed in white.
}
\label{intensities}
\end{figure}

We present in Fig. \ref{intensities} the images obtained in the two extreme photometric bands of
PRONAOS-SPM.
The angular resolution is 2$'$ in the 200 $\mic$ map, which corresponds to 1.3
parsec at the distance of M17, and 3.5$'$ in the 580 $\mic$ map (2.2 pc).
Due to the calibration uncertainty, the flux accuracy is 5 \% (1 $\sigma$) relative between bands (8 \% absolute).
We also present in Table \ref{fluxtab} the fluxes of the main identified regions,
integrated over a 3.5$'$ beam.

% FIGURE INTENSITY MAPS

% TABLE FLUXES
\begin{table}
\caption[]{Equatorial coordinates and fluxes (Jy) integrated over a 3.5$'$ area,
around the intensity peaks of the identified regions.
The absolute errors on the fluxes are 8 \% (1 $\sigma$).
These values take into account the fine color corrections due to the large
bandwidths of the SPM instrument.
}
\begin{flushleft}
\begin{tabular}{lllllll}

\hline
 & $\alpha_{1950}$ & $\delta_{1950}$ & F$_{\nu}$(Jy) & F$_{\nu}$(Jy) & F$_{\nu}$(Jy) & F$_{\nu}$(Jy) \\
 & (h,min,sec) & ($^o$,$'$) & 200 $\mic$ & 260 $\mic$ & 360 $\mic$ & 580 $\mic$ \\
\hline

M17 SW & 18 17 35 & -16 15 & 28000 & 15000 & 6600 & 1500 \\
\hline

M17 N & 18 17 45 & -16 03 & 4800 & 2500 & 1200 & 290 \\
\hline

Cloud A & 18 16 22 & -16 12 & 340 & 210 & 120 & 25 \\
\hline

Cloud B & 18 16 16 & -16 20 & 280 & 210 & 130 & 36 \\
\hline

Cloud C & 18 16 29 & -16 28 & 260 & 160 & 98 & 31 \\
\hline

Cloud D & 18 15 34 & -16 11 & 240 & 210 & 140 & 44 \\
\hline

\label{fluxtab}
\end{tabular}
\end{flushleft}
\end{table}

The maps in Fig. \ref{intensities} exhibit the very high intensity contrasts that exist in this
kind of
giant molecular complex.
The M17 Southwest (SW) and North (N) areas show up as the most intense dust
emission on our maps.
The M17 SW cloud reaches a peak intensity of 46000 \mjysr in band 1 (200
$\mic$ with a 2$'$ angular resolution) at the M17a source (see \cite{lada76}, \cite{wilson79} and \cite{gatley79}).
A small intense condensation is visible to the northeast of the most intense
area.
It shows up clearly in Fig. \ref{intensities} with the second highest contour in the 200 $\mic$ image,
and has a peak intensity of 9600 \mjysr at 200 $\mic$.
This condensation corresponds to the M17b source.
The M17 North condensation (also called M17c) reaches a peak intensity of 7800 \mjysr at 200 $\mic$.
The M17d source (\cite{wilson79}) is clearly visible to the east from M17b, and reaches a 200 $\mic$ peak intensity of 2000 \mjysr.

We have also mapped the area to the west of the M17 complex, in which we can
identify four condensations with weak intensities.
Cloud A (see Fig. \ref{intensities} for the names) is situated west of M17 SW and shows up as an
extended condensation reaching a 200 $\mic$ peak intensity of 630 \mjysr.
Cloud B is south of Cloud A and is linked to it by a bridge
clearly visible on the maps.
Cloud B reaches a 200 $\mic$ peak intensity of 440 \mjysr.
Another extended condensation, that we call Cloud C, is visible south of Cloud B.
It has a peak intensity of 470 \mjysr.
Cloud D is a condensation situated to the northwest of the maps,
linked to Cloud B through a filament, inside which a
small (2$'$) condensation is visible.
Cloud D has a 200 $\mic$ peak intensity of 470 \mjysr.

It is worth mentioning that these weak-intensity clouds are poorly visible in
the IRAS 100 $\mic$ survey.
As we can see in Fig. \ref{intensities}, the PRONAOS 580 $\mic$ map looks smoother than the
200 $\mic$ one, exhibiting less intensity contrast between regions.
As we shall discuss later, this is due to a trend of the intense areas to have
warmer dust than the faint ones.
This induces a stronger relative intensity of the weak-intensity areas in long-wavelength bands,
therefore reducing the contrast between the regions in the long-wavelength maps.

\section{Analysis\label{ana}}

\subsection{Dust temperatures and spectral indices\label{fit}}

\subsubsection{Derivation}

We assume that the emission of the grains is characterized by the modified blackbody law:

\begin{equation}
\inufit(\lambda,T,\beta) = C \; \bnu(\lambda,T) \; \lambda^{-\beta}
\end{equation}

where $\lambda$ is the wavelength, C a constant, T the temperature of the grains, $\beta$ the spectral
index and $\bnu$ the Planck function.
The fit procedure is described in detail in Dupac \etal (2001).
In this work, we included iteratively the color correction process due to the SPM bandwidths.
Assuming optically thin emission, we can derive
the optical depth $\tau_\nu={I_\nu\over{B_\nu}}=C \; \lambda^{-\beta}$.
The relative error on $\tau_\nu$ is ${\Delta I_{\nu}\over I_{\nu}} + {h \nu \over kT^2}\Delta T$.

\begin{figure}[]
\includegraphics{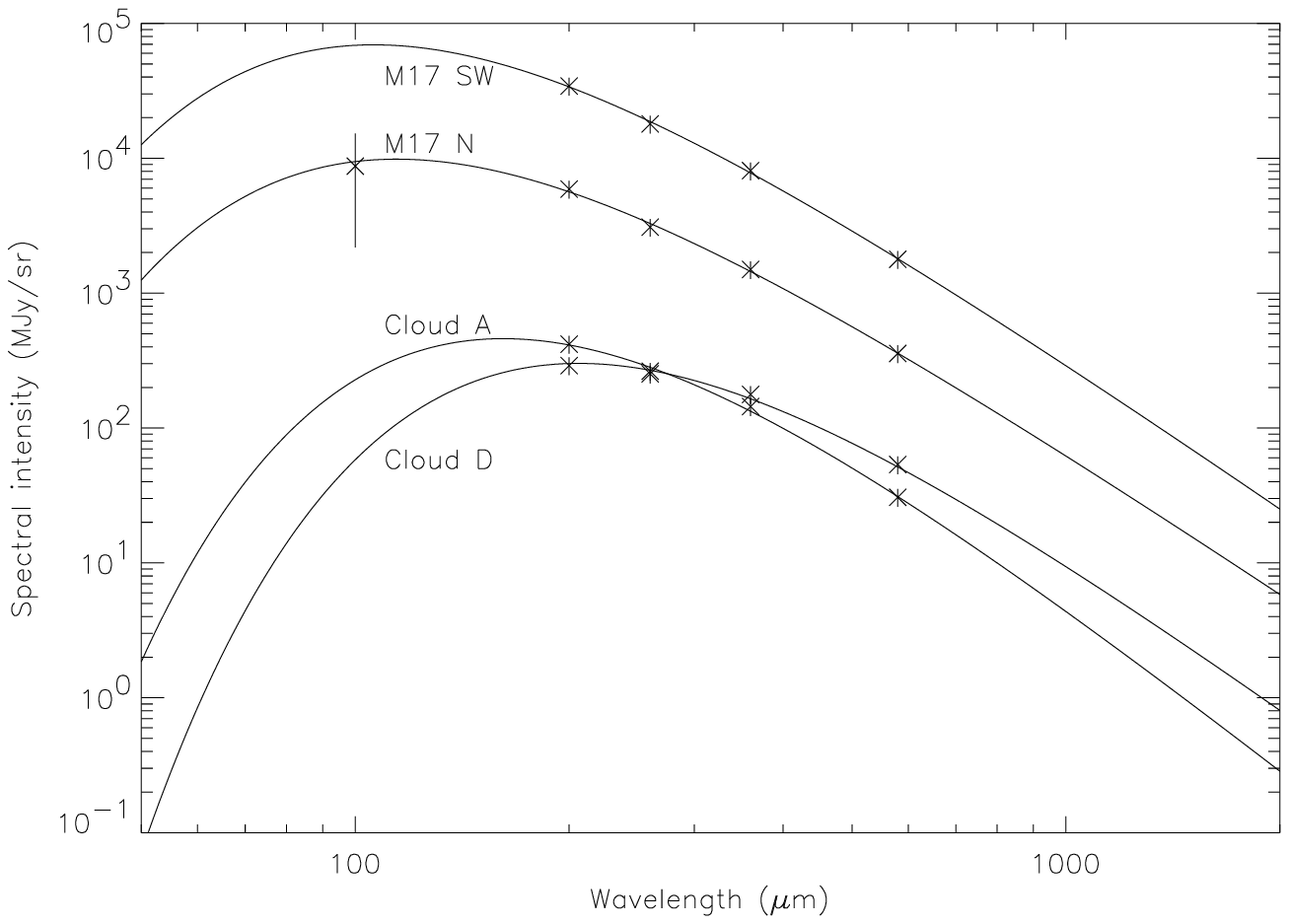}
\caption[]{Spectra with 3 $\sigma$ error bars.
The PRONAOS error bars are the intercalibration errors.
The 100 $\mic$ point is from the IRAS survey.
The drawn lines are the result of the fits (modified black body).}
\label{spectra}
\end{figure}

\begin{figure}[]
\includegraphics[scale=1.3]{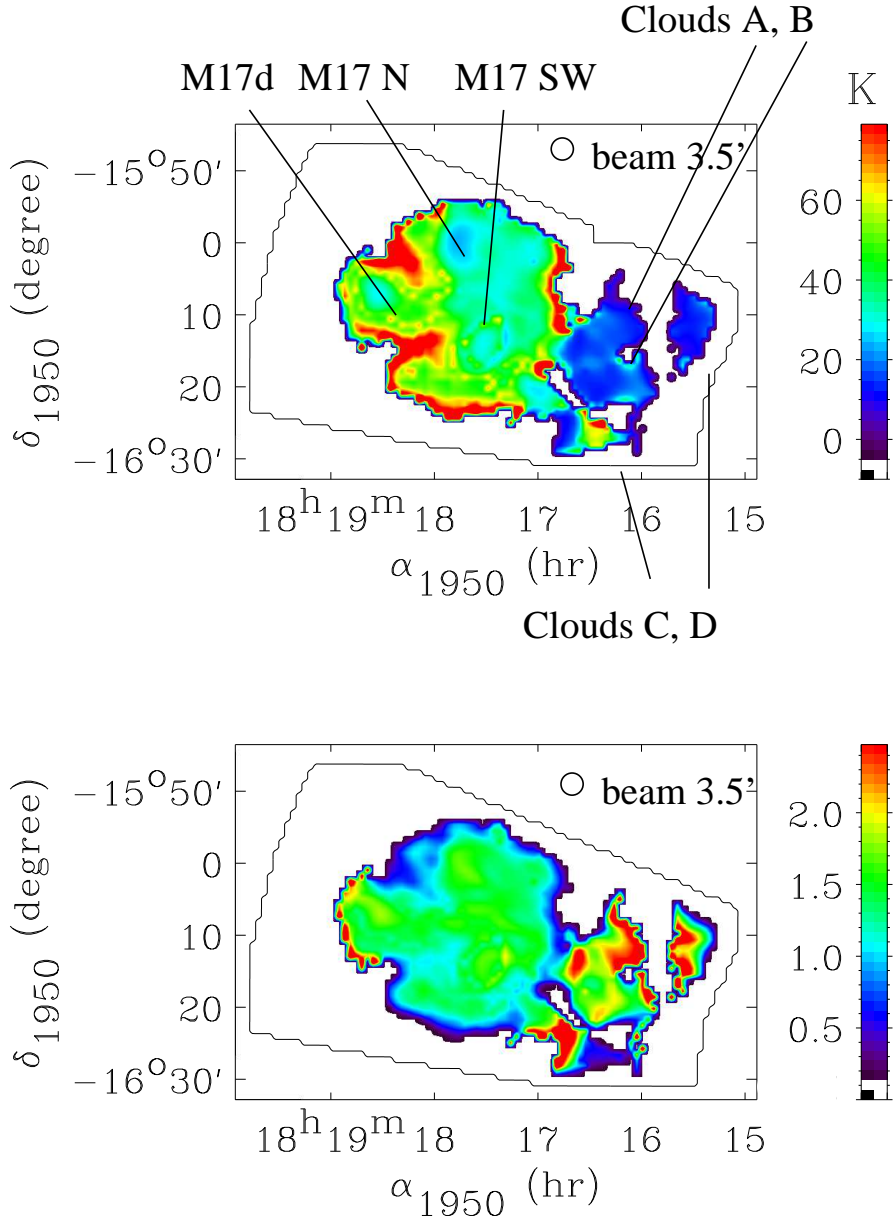}
\caption[]{Maps of the dust temperature in Kelvin (top) and the spectral index
  (bottom), made by fitting the modified blackbody to the PRONAOS spectra,
  with the IRAS 100 $\mic$ and 60 $\mic$ data for some regions.
The angular resolution is 3.5$'$ in each map.
The external contour shows the limit of the observed area.
The white regions inside this contour are too noisy to obtain an estimation of
  the temperature and the spectral index.
}
\label{tbeta}
\end{figure}

We show the temperature and spectral index maps derived this way in Fig. \ref{tbeta}.
These have been made without assuming anything about the temperature or the
spectral index.
We used the IRAS 100 $\mic$ data for the determination of the temperature and
the spectral index of a large part of the map (for this we assumed an intercalibration error
between IRAS and PRONAOS data of 25 \%), but neither for M17 SW (saturated) nor for the faint
clouds A, B, C, D to the west, for which the IRAS 100 $\mic$ data seem too noisy.
In the areas where the fit was difficult because of the high temperatures
estimated ($>$ 70 K), we used the IRAS 60 $\mic$ data too.
This is justified by considering that the thermal continuum of big grains
dominates the 60 $\mic$ emission for such high temperatures, whereas
it is likely not to be the case for lower temperatures for which the very small
grains might dominate the 60 $\mic$ emission.
With this process, the fit is performed in each pixel where the signal to
noise ratio is reasonable.
However, some observed areas are too noisy to obtain an estimation of the
temperature and the spectral index.
These areas are displayed in white on the maps in Fig. \ref{tbeta}.
We present spectra of some observed M17 regions in Fig. \ref{spectra}.

% FIGURE

% FIGURE

% TABLE TEMP-INDEX
\begin{table}
\caption[]{Temperature, spectral index and optical depth of the intensity peaks of the
identified regions.
The error bars are given for the 68 \% confidence interval.}

\begin{flushleft}
\begin{tabular}{lllll}
\hline
   & T (K) & $\beta$ & $\tau_{\nu} \centerdot 10^3$ & $\tau_{\nu} \centerdot 10^3$\\
   & & & 200 $\mic$ & 580 $\mic$\\

\hline

M17 SW & 29 $\pm$ 8 & 1.7 $\pm$ 0.3 & 80 $\pm$ 60 & 12 $\pm$ 3 \\
\hline

M17 N & 28 $\pm$ 3 & 1.6 $\pm$ 0.2 & 15 $\pm$ 6 & 2.6 $\pm$ 0.4 \\
\hline

Cloud A & 17 $\pm$ 3 & 2.3 $\pm$ 0.3 & 6 $\pm$ 4 & 0.5 $\pm$ 0.1 \\
\hline

Cloud B & 17 $\pm$ 3 & 1.7 $\pm$ 0.3 & 5 $\pm$ 3 & 0.7 $\pm$ 0.2 \\
\hline

Cloud C & 26 $\pm$ 4 & 1.2 $\pm$ 0.2 & 1 $\pm$ 0.5 & 0.3 $\pm$ 0.07 \\
\hline

Cloud D & 14 $\pm$ 2 & 1.9 $\pm$ 0.3 & 9 $\pm$ 6 & 1.3 $\pm$ 0.3 \\
\hline

\label{resul}
\end{tabular}
\end{flushleft}
\end{table}

\subsubsection{Variations of the temperature and the spectral index}

The maps in Fig. \ref{tbeta} exhibit interesting features: first, it is evident
comparing the maps that an anticorrelation exists between the
temperature and the spectral index.
We shall quantify this later.
In the major part of the maps, the temperature ranges from 10 K to 80 K, while
the spectral index exhibits also large spatial variations from 1 to
more than 2.5.
Temperatures above 80 K (around 100 K or higher) can be found in a few pixels near the ionization front to the east.
Large scale features on these maps are clearly visible, showing how the temperature increases from the west to the east.
The weak intensity areas to the west are significantly colder than the intense
M17 area.
We present in Table \ref{resul} the temperature, spectral index and optical depth for
the intensity peaks of the identified regions (Fig. \ref{intensities}).
The M17 SW region shows a
temperature of 29 K ($\pm$ 8) at the intensity peak.
For this area, we derive a spectral index of 1.7 ($\pm$ 0.3).
However, since the IRAS data are saturated in this area, the fit is quite
uncertain and the temperature could be higher.
The M17 N condensation has a temperature of 28 K, and a spectral index of 1.6.
The temperature in the M17 complex is about 30-50 K outside the M17 SW and M17 N
condensations, and the spectral index varies between 1 and 1.5.
On the edges of the complex, one can see the warm areas near the ionized regions (east and north-west), where
the temperature can be higher than 80 K, and the spectral index between 0.7
and 1.1.
The clouds A, B, and D in the western part of the maps show low temperatures
(14-17 K) and high spectral indices (1.7-2.3).
Some areas in these clouds even exhibit very low temperatures down to 10 K.

\subsubsection{Anticorrelation between the temperature and the spectral index}

\begin{figure}[]
\includegraphics{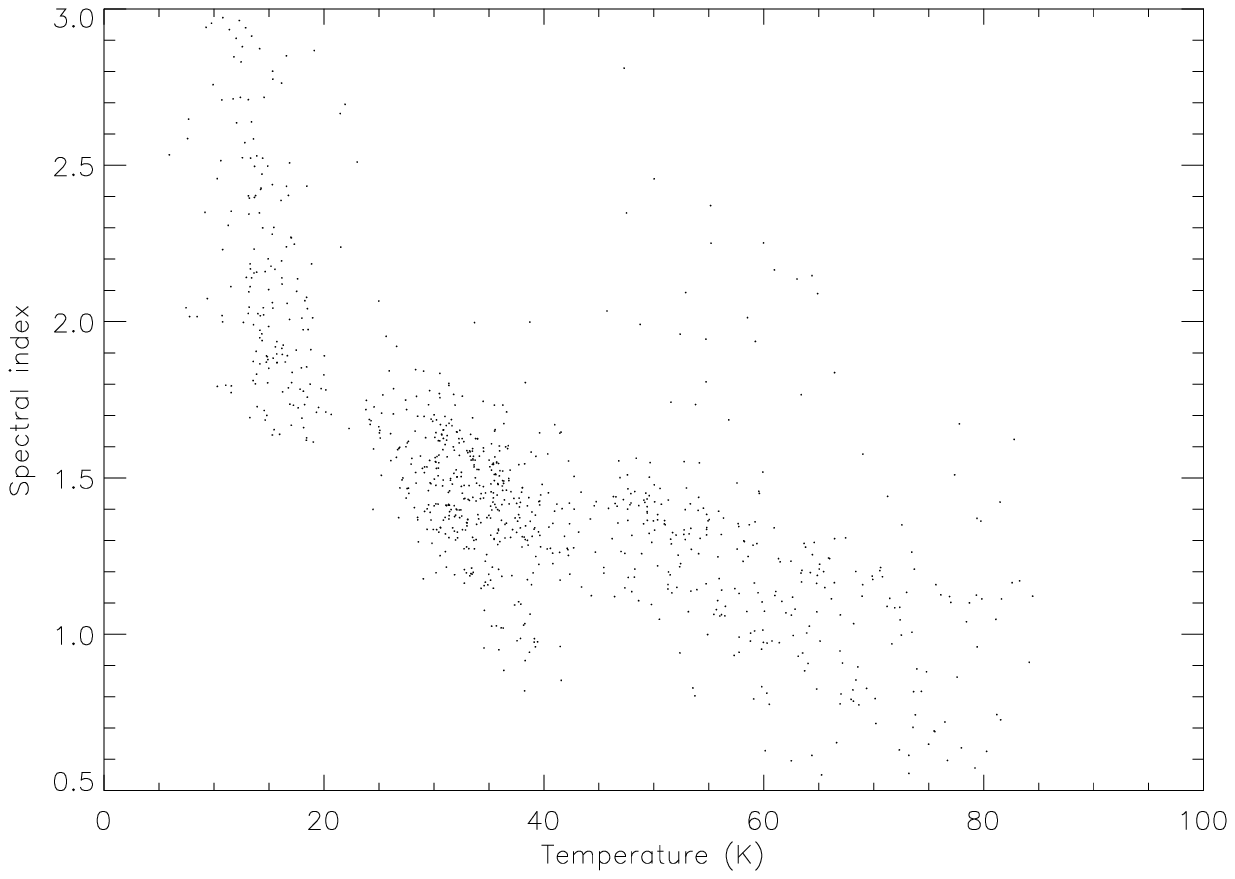}
\caption[]{Distribution in the (T,$\beta$) space of the (T,$\beta$) pairs
  derived with both errors on T and $\beta$ lower than 20 \%.
}
\label{maptb}
\end{figure}

As we can see in Fig. \ref{tbeta}, it seems that an anticorrelation exists
between the temperature and the spectral index.
This effect was shown for the first time in observations by Dupac \etal (2001) in
Orion.
If we take the five intensity peaks presented in Table \ref{resul}, the correlation coefficient is -0.58, which means a significant anticorrelation.
However, it is not highly statistically significant, given the small number of points.
Therefore, we have analyzed the correlation between the temperature and the spectral index in the whole map.
We present in Fig. \ref{maptb} a plot of the distribution of the pixels in Fig. \ref{tbeta} in
the (T,$\beta$) space.

% FIG.

One can see the banana shape of this distribution in Fig. \ref{maptb}, which clearly shows the anticorrelation between both parameters.
In particular, very few pixels can be found in the data with T $>$ 40 K and $\beta$ $>$ 1.5,
and no pixel can be found with T $<$ 20 K and $\beta$ $<$ 1.6.
If we consider all the (T,$\beta$) pairs with T between 10 K and 80 K, and
$\beta$ between 1 and 2.5, then the correlation coefficient is -0.56.
Among these 1079 pixels, 1071 have both temperature and spectral index with relative
errors less than 50 \%, and 801 with relative errors less than 20 \%.
The correlation coefficient computed from these 801 well fitted pixels is -0.64.
If we consider only the pixels with relative errors less than 10 \% (87
pixels), then the correlation is -0.74.
This increasing of the anticorrelation with the restriction on the errors is
evidence of the reality of this effect.
Also, pixels with high spectral indices have narrow error bars on the spectral index: there are 70 pixels for 2.3 $< \beta <$ 3, with an average
relative error on $\beta$ of 12 \% - the relative errors are quite uniform
for these points.
Pixels with low spectral indices also show quite narrow error bars: there are 482 pixels for 0.7 $< \beta <$ 1.3, with an average relative error on $\beta$ of 16 \%.
Therefore, we can be confident of this anticorrelation effect.
Nevertheless, to better understand it, we
performed simulations of PRONAOS data from the same number (801) of random
(T,$\beta$) pairs in the same (T,$\beta$) range, and fitted them the same way as real data.
We repeated this processing with several distributions of T-$\beta$ pairs without
intrinsic correlation, with and without 100 $\mic$ simulated data, and
obtained correlation coefficients on the fitted parameters usually between 0.0
and -0.2, and never above (in absolute) -0.3.
Therefore, the fit procedure
itself involves some degree of anticorrelation between the derived temperature
and spectral index, as Dupac \etal (2001) had already investigated.
There is indeed some degeneracy between both parameters,
but this correlation coefficient which we find on simulations is clearly not enough
to explain the correlation found on the data (-0.64).
Thus we have to conclude that this anticorrelation effect has to be either an intrinsic physical property of
the grains, or at least a property of the observed dust, when integrated on
the beam and the ISM column.
The same effect has been shown in Orion (\cite{dupac01}).

To interpret this anticorrelation, we can investigate what the meaning is of such spectral index variations.
Mixtures of dust components with different temperatures can be put forward: this clearly cannot explain high indices, but could explain low indices around 1.
Indeed, warm dust can be associated in the same line of sight with cold
dust components which could enhance the high-wavelength intensity, and
therefore decrease the spectral index from 2 to around 1.
However, this would imply a large amount of cold dust (see the
investigations of \cite{ristorcelli98} and \cite{dupac01}).
In this article, we have performed other studies.
We simulate spectra with low spectral indices and various warm temperatures.
We then try to fit these spectra with two dust components having both a ``standard'' spectral index equal to 2, the warm component having the same 100 $\mic$ emission and the same temperature as the single component spectrum.
Thus we have equality between both optical depths ($C \; \lambda^{-\beta}$) at 100 $\mic$, and therefore, given the simple column density model that we present in Section \ref{col}, the warm dust component has the same column density as the single dust component.
In this way, we fit the short-wavelengths part of the original spectrum by the warm component, but the discrepancy between the original spectral index and the standard one ($\beta$ = 2) induces a lack of emissivity of the warm component compared to the original spectrum at long wavelengths.
Therefore, we introduce a colder dust component with the same standard spectral index $\beta$=2, which we add to the warm component to fit the spectrum.
Then we can investigate what amount of cold dust is necessary to explain the observed low spectral indices, by computing the column densities from our simple model (see Section \ref{col} and Dupac \etal 2001).
Indeed, with two dust components with the same standard spectral index, the column density ratio between both is simply the ratio of the parameters C.

% TABLE: TWO COMPONENT EMISSION
\begin{table}
\caption[]{Summary of the investigations made concerning the two-dust component assumption.
From left to right: the original spectrum temperature, the original spectral index, the temperature of the cold component, which has a spectral index $\beta$=2, and the mass ratio derived between the cold and warm components.}

\begin{center}
\begin{tabular}{llll}
\hline
T (K) & $\beta$ & T$_{cold}$ (K) & Mass ratio \\
\hline
50 & 1 & 10 & 100 \\
\hline
70 & 1 & 10 & 200\\
\hline
60 & 1 & 20 & 20\\
\hline
60 & 1 & 40 & no fit\\
\hline
50 & 1.5 & 10 & 50\\
\hline
50 & 1.5 & 20 & 5\\
\hline
50 & 1.5 & 30 & 3, bad fit\\
\hline
40 & 1.7 & 10 & 20\\
\hline
\label{twocomp}
\end{tabular}
\end{center}
\end{table}

We present in Table \ref{twocomp} a summary of these investigations.
It appears that low indices (around 1) would imply very large amounts of cold dust in the line of sight.
For example, to fit this way a spectrum with T = 50 K, $\beta$ = 1, with a 10 K component, a cold dust column density 100 times larger than the warm dust column density is necessary.
We have said that the column density of the warm component is the same as the single dust-method one.
As seen in Table \ref{masses}, the column density estimates, assuming one dust component, seem to have the right order of magnitude, compared for example to the estimates from the CO.
It is extremely unlikely that there are such huge amounts of cold dust (much larger than the warm dust and the CO estimates) in each line of sight where we observe low indices.
If one uses intermediate temperatures for the cold component, then it is no longer possible to match the spectrum at long wavelengths, even for $\beta$=1.5.
We show by these investigations that even for intermediate spectral indices, it is unlikely that they be explained only by temperature mixtures with the same standard spectral index.

Moreover, the shape of the pixel distribution in the (T, $\beta$) space is very difficult to explain with only the argument of temperature mixtures.
It would suggest that the warmer the (warm) dust
component, the more massive the cold component, because the spectral
index is further reduced (to 1) at high temperatures.
This can certainly not be a general rule, and this shows that the temperature mixture assumption is insufficient to explain the variations found on the spectral index and its inverse dependence on the temperature.

Laboratory experiments (see
\cite{agladze96} and \cite{mennella98}) showed this anticorrelation effect on
grains for temperatures down to 10 K.
Agladze \etal (1996) measured absorption spectra of crystalline and amorphous
grains between 0.7 and 2.9 mm wavelength. They deduced an anticorrelation
between the power-law index $\beta$ and the temperature in the temperature range
10-25 K, and attributed it to two-level tunnelling processes.
The measures of Agladze \etal (1996) are insufficient to justify
our observation in the submillimeter spectral range, because absorption can be
very different in the millimeter range.
Mennella \etal (1998) measured the absorption coefficient of cosmic dust grain
analogues, crystalline and amorphous, between 20 $\mic$ and 2 mm
wavelength, in the temperature range 24-295 K. They deduced an
anticorrelation between T and $\beta$, and attributed it to two-phonon
difference processes.
In our observations, we observe this effect down to about 10 K (Fig. \ref{maptb}), thus
we would need laboratory results on these low temperatures in the
submillimeter range to fully understand the observations.
Other causes can make the spectral index vary, such as the composition and size of the
grains.

We can conclude from this temperature-spectral index analysis that the
anticorrelation discovered by Dupac \etal (2001) in Orion is shown
in M17 too, another high-mass star-forming complex of our Galaxy.
Our investigations
%like the total absence of warm dust ($>$ 40 K) with high indices ($>$ 1.5)
make us rather believe in a fundamental property of the grains to explain this effect.

\subsection{Column densities\label{col}}

We are interested in estimating the column densities and masses of the studied
region.
For this we follow the simple model described in Dupac \etal (2001), which uses
the dust 100 $\mic$ opacity from D\'esert \etal (1990).
We derive column densities that we present in Table \ref{masses}.
To obtain an independent estimation of the column density, we use observations
of the rotational transition $J = 1-0$ of $^{13}CO$, made by Wilson \etal
(1999).
Their map covers the M17 cloud but not the fainter area to the west.
Assuming local thermal equilibrium and optically thin emission, the
$^{13}CO$ column density can be expressed as:

\begin{equation}
N_{^{13}CO} = 2.6 \: 10^{14} \; {W_{^{13}CO} \over (1-e^{-5.3/T})}
\end{equation}

(see e.g. \cite{rohlfs00})
Then, assuming a ${H_2 \over ^{13}CO}$ number ratio of 4.6 $10^5$
(\cite{rohlfs00}), a mean molecular weight of 2.36 $m_H$ (see e.g.
\cite{elmegreen79}), and
a temperature of the gas of 30 K, we derive a column density to $^{13}CO$ line
ratio of 17 $10^{20}$ protons $cm^{-2} (K.km/s)^{-1}$.
Using this coefficient, we derive column densities that we present in Table \ref{masses},
in order to compare to the PRONAOS + D\'esert model estimation.

% TABLE 3
\begin{table}
\caption[]{Column densities estimated from the PRONAOS+D\'esert method, and from the $^{13}$CO
  data of Wilson \etal (1999), masses from PRONAOS+D\'esert,
  PRONAOS+Ossenkopf, Jeans masses.
The error bars on the PRONAOS+D\'esert column densities are given for the 68 \% confidence interval.}

\begin{flushleft}
\begin{tabular}{llllll}
\hline
&N$_{H}$ \tiny{PRONAOS+D\'es.} & N$_{H}$ \tiny{$^{13}$CO} & Mass
\tiny{PRONAOS+D\'es.} & Mass \tiny{PRONAOS+Oss.} & Jeans mass \\

&$10^{20} H cm^{-2}$ & $10^{20} H cm^{-2}$ & \msol & \msol & \msol \\
\hline

M17 SW & 4100 $\pm 3600$ & 1200 & 16000 & --- & --- \\
\hline

M17 N & 730 $\pm 340$ & 460 & 4600 & --- & --- \\
\hline

Cloud A & 450 $\pm 380$ & --- & 2000 & 730 & 70\\
\hline

Cloud B & 250 $\pm 210$ & --- & 2400 & 870 & 100\\
\hline

Cloud C & 38 $\pm 25$ & --- & 350 & --- & 150 \\
\hline

Cloud D & 630 $\pm 530$ & --- & 2100 & 760 & 100\\
\hline

\label{masses}
\end{tabular}
\end{flushleft}
\end{table}

The agreement is not very good in M17 SW between PRONAOS + D\'esert
model and $^{13}$CO, though it is consistent with the error bar.
It is not unlikely that this is due to optical thickness of the $^{13}$CO 1-0
emission.
The M17 N region shows a better agreement.
If we trust our PRONAOS + D\'esert model estimation (which seems quite
speculative in view of the error bars), we can constrain the column density to $^{13}CO$ line
ratio in M17, which we find to range roughly from 15 $10^{20}$ to 100 $10^{20}$
following the regions of M17 (in units of protons $cm^{-2} (K.km/s)^{-1}$).
Since the ratio corresponding to optically thin $^{13}CO$ emission is 17
$10^{20}$, and that the most intense areas have the highest coefficient, it
is possible to conclude that the disagreement between PRONAOS and
$^{13}CO$ column density estimations of M17 SW may be due (at least in a
part) to optical thickness of the $^{13}CO$ 1-0 line.
Also, uncertainties in the ${H_2 \over ^{13}CO}$ ratio may arise for high
column densities.
Moreover, the excitation
temperature can be higher than 30 K in M17 SW, although we measure
it to be around 30 K.
If we take an excitation temperature of 60 K, then the column
density estimated from the $^{13}CO$ emission is 2300 $10^{20}$ H $cm^{-2}$
for M17 SW (at the intensity peak).
If now we look at the sources of uncertainties in the PRONAOS + D\'esert model
estimation - apart from the error bars being large, we can point out that
the 100 $\mic$ opacity we used might overestimate the column density,
especially in cold clouds where could take place some special effects like
formation of molecular ice mantles on the grains and coagulation of grains.
For instance, these processes are taken into account by the protostellar core
model of Ossenkopf \& Henning (1994), for which the 100 $\mic$ opacity is 1
$cm^{2}/g$.
By making the same analysis as with the model of D\'esert \etal (1990), using
the Ossenkopf \& Henning opacity,
we derive column densities 2.8 times lower.
This correction could be true for cold clouds like A, B and D.

Finally, we computed masses of regions in this giant molecular complex.
For this, we integrated the column density found from our submillimeter
measurement over the area of the clouds that we observe in the maps, assuming a
distance of 2200 pc (\cite{chini80}).
By this way, we derive a total ISM mass of the M17 complex
(without Clouds A, B, C, D to the west) of 31000 \msol, in which M17 SW
weights 16000 M$_{\odot}$ and M17 N 4600 \msol.
We derive also mass estimates of the clouds to the west, which we present in
Table \ref{masses}.
For the cold clouds A, B and D, it is likely that the masses are smaller than
that: if we trust the Ossenkopf \& Henning (1994) model, then we derive masses
of about 800 M$_{\odot}$ for each.
For the clouds separated from the molecular complex (A, B, C, D), we present
also in Table \ref{masses} the Jeans masses, as an indicator of the
gravitational stability of these condensations (see \cite{dupac01} for details).
There are of course large uncertainties
in both the clump's measured mass and the non-thermal sources of internal
energy, such as magnetic field and turbulence, which can help to balance the
gravitational energy.
Nevertheless, we observe that Clouds A, B and D are possibly gravitationally unstable,
especially Cloud A, which might be compressed by the external pressure due to the
ionized area to the north.

\section{Conclusion}

We observed a large (50$'$ x 30$'$, $\approx$ 30 pc x 20 pc) area in and around
the M17 molecular complex.
We showed extended faint clumps (A, B, C, D) in the outskirts of the active star-forming area.
Our study shows a large distribution of temperatures and spectral indices:
the temperature varies roughly from 10 K to 100 K,
and the spectral index from 1 to 2.5.
The statistical analysis of the temperature and spectral index spatial
distribution shows the anticorrelation between these two
parameters.
Indeed, we observe cold dust (10-20 K) with high indices (around 2), and warm
dust ($>$ 20 K) with low indices (around 1 - 1.5), but we could not
significantly find any cold place with low indices nor warm place with high indices.
The investigations which we made to match the observed spectra, with two dust components having a standard spectral index, imply very large and unlikely amounts of cold dust, therefore we rather support a fundamental explanation for this effect.
Laboratory measurements showed this anticorrelation in the submillimeter domain on
grains for temperatures down to 25 K, but we would need laboratory results on
temperatures down to 10 K in the submillimeter range to fully understand our observations.
This anticorrelation effect was also shown in Orion (\cite{dupac01}), and in
other PRONAOS observations that are still being analyzed.

We estimated the column densities and masses of the observed regions by simply
modelling the thermal emission of the grains from D\'esert \etal (1990) or
Ossenkopf \& Henning (1994).
We derived a total mass of the M17 cloud of 31000 \msol.
Of course, the error bars are large and difficult to estimate properly,
but there is a clear trend for three cold clouds (A, B, D) to be
gravitationally unstable, and therefore to be pre-stellar candidates.

These observations could be sustained by sensitive continuum
observations in the millimeter domain, especially for
the faint clouds studied to the west of the M17 complex, in order to better
constrain the spectral index measurement of the cold dust.
Also, infrared observations at higher resolution could be useful to study the
structure of these cold clouds.

\section{Acknowledgements}
We thank very much Christine Wilson and her collaborators for having provided
us their $^{13}CO$ data.
We are indebted to the French space agency Centre National d'\'Etudes Spatiales
(CNES), which supported the PRONAOS project.
We are very grateful to the
PRONAOS technical teams at CNRS and CNES, and to the NASA-NSBF balloon-launching facilities group of Fort Sumner (New Mexico).

\end{document}